\documentclass[aps,prl,reprint,groupedaddress,showpacs,amsmath,amssymb,twocolumn,floatfix]{revtex4-1}
\usepackage{graphicx}
\usepackage{bm}
\usepackage{color}
\usepackage[normalem]{ulem}
\usepackage{amsmath}
\usepackage{tabularx}

\usepackage[colorlinks=true]{hyperref}  
\hypersetup{
    unicode=false,          
    pdftoolbar=true,        
    pdfmenubar=true,        
    pdffitwindow=false,     
    pdfstartview={FitH},    
    pdftitle={XXX},    
    pdfauthor={XXX et al.},     
    pdfsubject={},   
    pdfcreator={},   
    pdfproducer={}, 
    pdfkeywords={} {} {}, 
    pdfnewwindow=true,      
    colorlinks=true,       
    linkcolor=magenta, 
    citecolor=blue,        
    filecolor=magenta,      
    urlcolor=blue           
} 
\usepackage{cleveref}

\newcommand{\Rpara}{$R_{\parallel}$}
\newcommand{\Rperp}{$R_{\perp}$}

\newcommand{\tlg}{tTLG}
\newcommand{\WSe}{WSe$_2$}

\newcommand{\DRR}{$\Delta R / R_0$}

\newcommand{\sigbar}{$\bar{\sigma}$}
\newcommand{\dsig}{$\delta \sigma$}

\newcommand{\Vto}{$\Delta V_{\perp}^{2\omega}$}

\usepackage{pifont}

\usepackage{bbm}

\renewcommand{\vec}[1]{\boldsymbol{#1}}

\begin{document}

\title{Electronic anisotropy in magic-angle twisted trilayer graphene}

\author{Naiyuan J. Zhang$^{1}$}
\author{Yibang Wang$^{1}$}
\author{K. Watanabe$^{2}$}
\author{T. Taniguchi$^{3}$}
\author{Oskar Vafek$^{4,5}$}
\author{J.I.A. Li$^{1}$}
\email{jia\_li@brown.edu}

\affiliation{$^{1}$Department of Physics, Brown University, Providence, RI 02912, USA}
\affiliation{$^{2}$Research Center for Functional Materials, National Institute for Materials Science, 1-1 Namiki, Tsukuba 305-0044, Japan}
\affiliation{$^{3}$International Center for Materials Nanoarchitectonics,
National Institute for Materials Science,  1-1 Namiki, Tsukuba 305-0044, Japan}
\affiliation{$^{4}$Department of Physics, Florida State University, Tallahassee, FL 32306, USA}
\affiliation{$^{5}$ National High Magnetic Field Laboratory, Tallahassee, Florida, 32310, USA}

\date{\today}

\maketitle

\textbf{Due to its potential connection with nematicity, electronic anisotropy has been the subject of intense research effort on a wide variety of material platforms.
The emergence of spatial anisotropy not only offers a characterization of material properties of metallic phases, which cannot be accessed via conventional transport techniques, but it also provides a unique window into the interplay between Coulomb interaction and broken symmetry underlying the electronic order.
In this work, we utilize a new scheme  of angle-resolved transport measurement (ARTM) to characterize electron anisotropy in magic-angle twisted trilayer graphene.   By analyzing the dependence of spatial anisotropy on moir\'e band filling, temperature and twist angle, we establish the first experimental link between electron anisotropy and the cascade phenomenon, where Coulomb interaction drives a number of isospin transitions near commensurate band fillings  ~\cite{Cao2018a,Park2021flavour,Saito2021pomeranchuk,Rozen2020pomeranchuk}. Furthermore, we report the coexistence between electron anisotropy and a novel electronic order that breaks both parity and time reversal symmetry. Combined, the link between electron anisotropy, cascade phenomenon and $PT$-symmetry breaking sheds new light onto the nature of electronic order in magic-angle graphene moir\'e systems. }

\begin{figure*}
\includegraphics[width=1\linewidth]{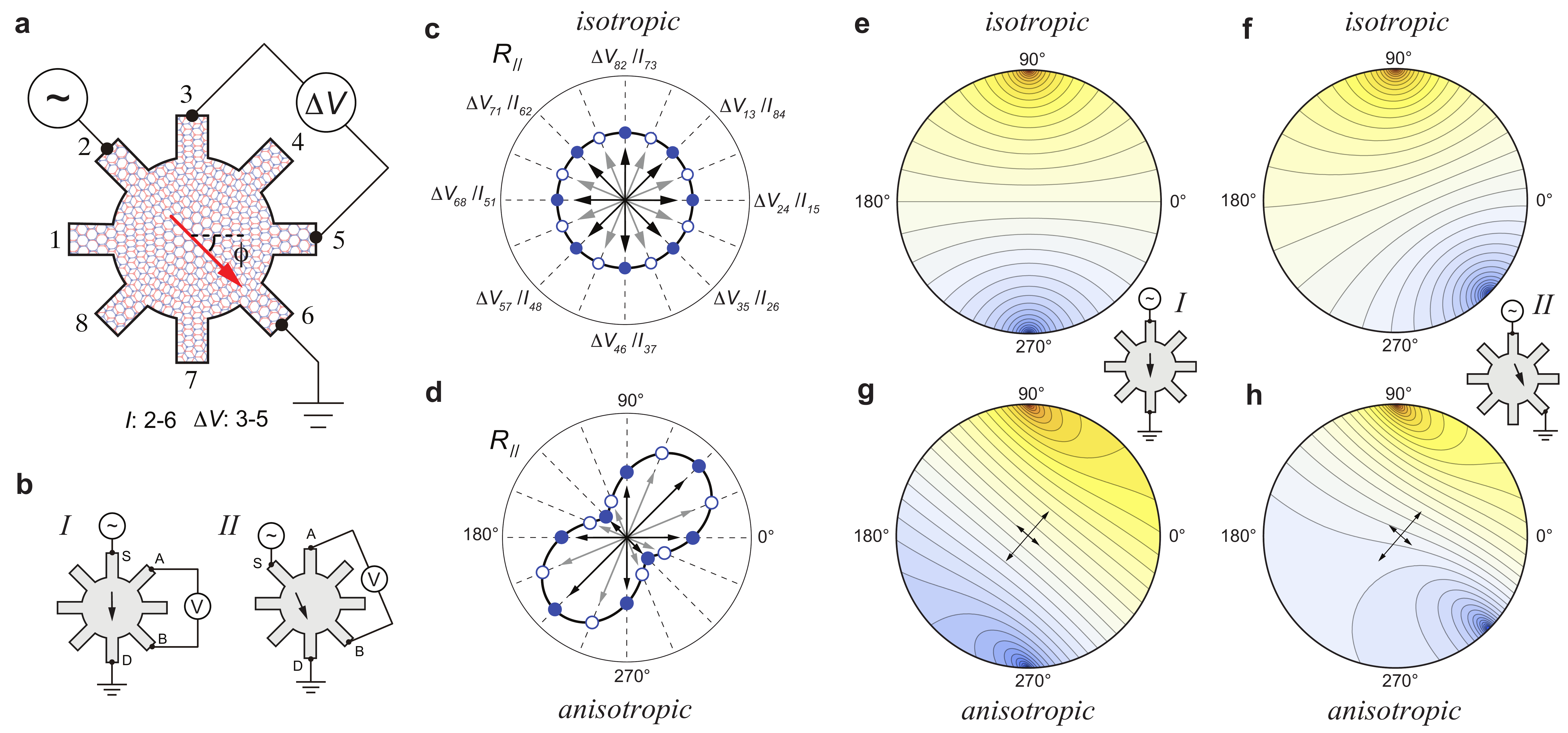}
\caption{\label{fig1}{\bf{The ``sunflower'' geometry.}} (a) Schematic showing  transport measurement setup on a sample with the ``sunflower'' geometry. Eight electrical contacts are labelled $1$ through $8$. In this setup, a DC current bias is applied to contact $2$ and $6$, which are also referred as the source and drain contacts.  Voltage drop $\Delta V$ is measured across two electrical contacts, $3$ and $5$.  (b) Schematic diagram showing two current bias configurations $I$ and $II$ that are used in this work. When voltage leads are aligned parallel with current flow direction, the ratio $\Delta V_{\parallel}/I (\phi)$ is comparable to the longitudinal resistance, which will be referred to as $R_{\parallel}(\phi)$. $\phi$ denotes the azimuthal direction of current flow. (c-d) Schematic diagram showing the angular dependence of $R_{\parallel}(\phi)$ for (c) an isotropic state and (d) an anisotropic state.  (e-h) Schematic showing the distribution of electrical potential across a uniform sample with different boundary conditions and conductivity tensors. Panel (e) and (g) are calculated with the boundary condition of configuration I, whereas panel (f) and (h) for configuration II. Panel (e) and (f) assumes an electronic state with an isotropic conductivity tensor, whereas (g) and (h) are for an anisotropic state.  }
\end{figure*}

Electronic nematic, a translationally invariant metallic phase that breaks the in-plane rotational symmetry of the underlying crystal lattice, is a hallmark of strongly correlated electronic systems ~\cite{Fradkin2010nematic,Oganesyan2001nematic,Kivelson1998nematic}. Spatial anisotropy in electronic states has been observed in a variety of material platforms, such as two-dimensional electron systems (2DES) at high magnetic fields ~\cite{Lilly1999stripe,Du1999stripe}, strontium ruthenate and cuprate materials ~\cite{Wu2017nematic,Wu2020nematic,Ando2002nematic,Hinkov2008nematic}. 
Recently, electron anisotropy has been reported in the superconducting and normal phases of graphene-based moir\'e systems ~\cite{Jiang2019STM,Choi2019STM,Kerelsky2019STM,Cao2020nematicity,Rubio2022nematic}.  Owing to the quenched electron kinectic energy,  Coulomb interaction plays a prominent role in determining the electronic order within the moir\'e flatband. This is reflected by a cascade of isospin transitions near integer band fillings, which lifts the spin and valley degeneracy and reconstructs the Fermi surface with well-defined isospin orders ~\cite{Park2021flavour,Zondiner2020cascade,Wong2020cascade,Kang2021cascades}. A number of theoretical works have recognized a possible connection between electron anisotropy and strong Coulomb interaction within the moir\'e band ~\cite{Liang2019tblg,Chichinadze2019nematic,Liu2021nematic,Parker2021nematic,Zhang2022nematic,Wagner2022strain,Samajdar2021nematic,SboyChakov2020nematic,Fernandes2020nematic,Kang2020nematic}.
 However,  experimental evidence directly demonstrating this link has remained elusive.

The effort to understand the interplay between Coulomb interaction, isospin order and spatial anisotropy is complicated by the large moir\'e wavelength of graphene-based moir\'e systems. A recent calculation of single-particle band structure pointed out that the influence of lattice distortion is amplified by the large moir\'e wavelength in twisted bilayer graphene and that even a small amount of heterostrain, on the order of $0.2\%$, could induce prominent electron anisotropy. Most strikingly, strain-induced anisotropy is shown to exhibit doping-dependence in both the magnitude and the orientation of the director axis  ~\cite{Wang2022strain}. Therefore, the observation of doping dependence in the orientation of anisotropy director is insufficient to isolate the role of Coulomb interaction in inducing electron anisotropy ~\cite{Wu2017nematic,Choi2019STM,Rubio2022nematic}.
The large moir\'e wavelength also gives rise to an abundance of inhomogeneity in the spatial distribution of the twist angle ~\cite{Uri2020mapping,Mcgilly2020moire}, which provides additional challenges for experimental efforts to characterize the nature of electron anisotropy. 
In this work, we utilize a new scheme of angle-resolved transport measurement (ARTM) to simultaneously extract the conductivity matrix and characterize the spatial uniformity of the electronic state in magic-angle twisted trilayer graphene. Not only does ARTM demonstrate a direct link between electron anisotropy and the cascade phenomenon, but it also provides a new route for unraveling the nature of electronic orders across the moir\'e flatband.

The ARTM is enabled by the ``sunflower'' device geometry, as shown in Fig.~\ref{fig1}a.
The circular part of the sample is designed with a diameter of $\sim 2 \mu$m to minimize the influence of twist angle inhomogeneity. Electrical contacts are made to eight ``petals'', which are labelled $1$ through $8$ (Fig.~\ref{fig1}a). A measurement in the ``sunflower'' geometry is carried out by applying current bias to a pair of contacts while measuring the voltage difference across a different pair. For simplicity, we use $\Delta V_{35}/I_{26}$ to denote the measurement configuration shown in Fig.~\ref{fig1}a, where current flows from contact $2$ to $6$ and voltage difference is measured between contacts $3$ and $5$. The ``sunflower'' geometry allows for $840$ independent measurement configurations: allowing for non-reciprocity, there are $56$ ways to pick the source and the drain and for each choice there are $15$ voltage lead pairs. 
In this work, we focus on two types of measurement configurations, as shown in Fig.~\ref{fig1}b, In configuration I, current bias is applied to contact $i$ and $i+4$, whereas current bias in configuration II is applied to contact $i$ and $i+3$. For each configuration, $\Delta V_{\parallel}$ and $\Delta V_{\perp}$ is defined as the voltage difference across contacts that are aligned parallel and perpendicular to the direction of current flow, respectively (for definitions, see Fig.~\ref{figMeasurement}). For simplicity, we will refer to $\Delta V_{\parallel}/I$ and $\Delta V_{\perp}/I$ as \Rpara\ and \Rperp, which are directly comparable to the longitudinal and transverse resistance of the sample. As $i$ varies through $1$ to $8$, configuration I and II allow us to measure \Rpara\ and \Rperp\ with $16$ azimuthal directions of current flow, with an angular resolution of $\phi = 22.5^{\circ}$.  As shown in Fig.~\ref{fig1}c-d, the evolution of \Rpara\ with varying $\phi$ offers a direct identification for electronic anisotropy.

\begin{figure*}
\includegraphics[width=0.95\linewidth]{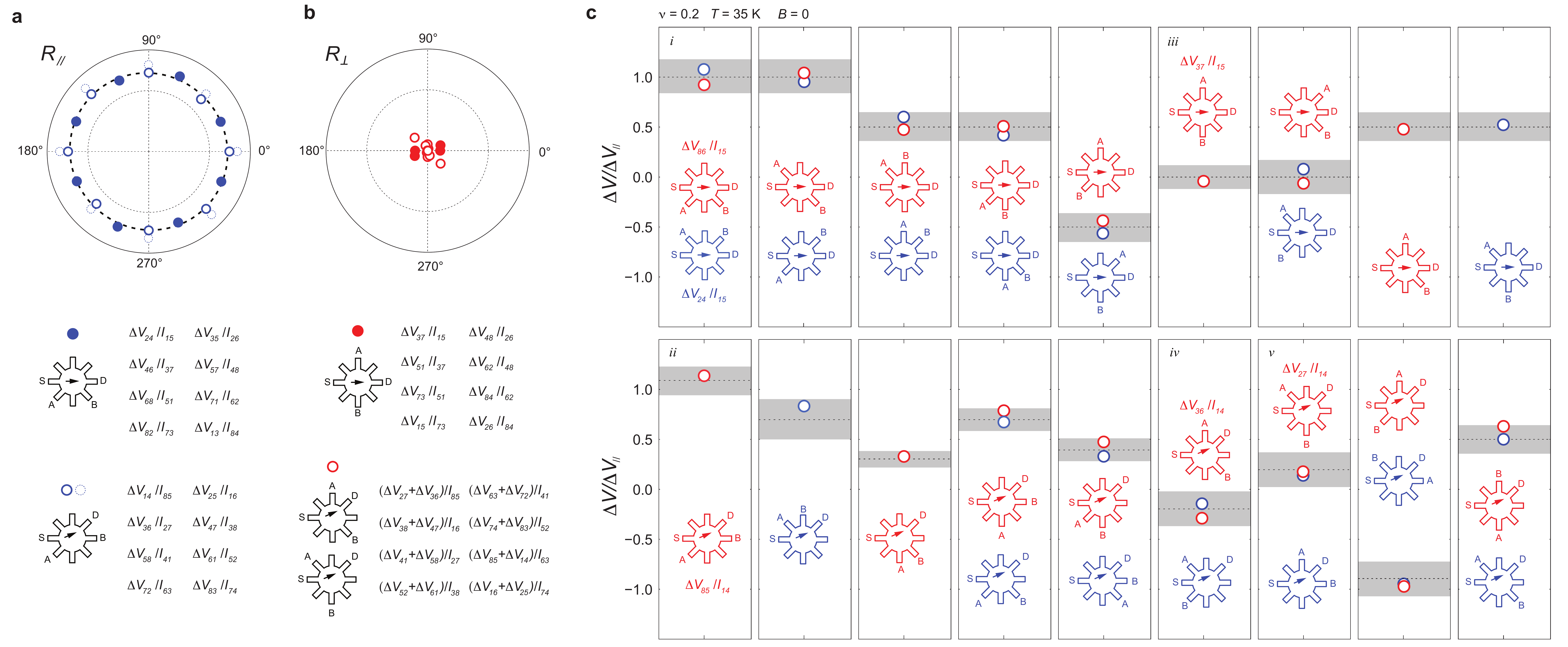}
\caption{\label{fig2} {\bf{Angle-resolved transport response of an isotropic state at $T =35$ K.} }  (a) \Rpara\ and  (b) \Rperp\ as a function of azimuth direction $\phi$ measured at $T = 35$ K and $\nu=0.2$. The measurement configuration for each data point is labeled in the legend. Solid (open) circles are measured with configuration I (II). The dashed open circles are measurement value, whereas the open circles is the corrected value accounting for the different geometry between configurations I and II. The measurement configurations for $R_{\parallel}=\Delta V_{\parallel}/I$ and $R_{\perp}=\Delta V_{\perp}/I$ are labeled to the bottom. (c) Comparison between $\Delta V$ measured from $30$ different configurations and the expected voltage difference for each configuration calculated using an isotropic conductivity tensor. The black dashed line denote the expected value.  
The plotted value across all panels is renormalized by the expected value of $\Delta V_{\parallel}$ from the configuration I.   The width of the stripes denote the expected error arising from the non-zero width of voltage leads. Panel \emph{i-ii} (panel \emph{iii-v}) correspond to $\Delta V_{\parallel}$ ($\Delta V_{\perp}$) measured with configuration I and II.
}
\end{figure*}

\begin{figure*}
\includegraphics[width=0.95\linewidth]{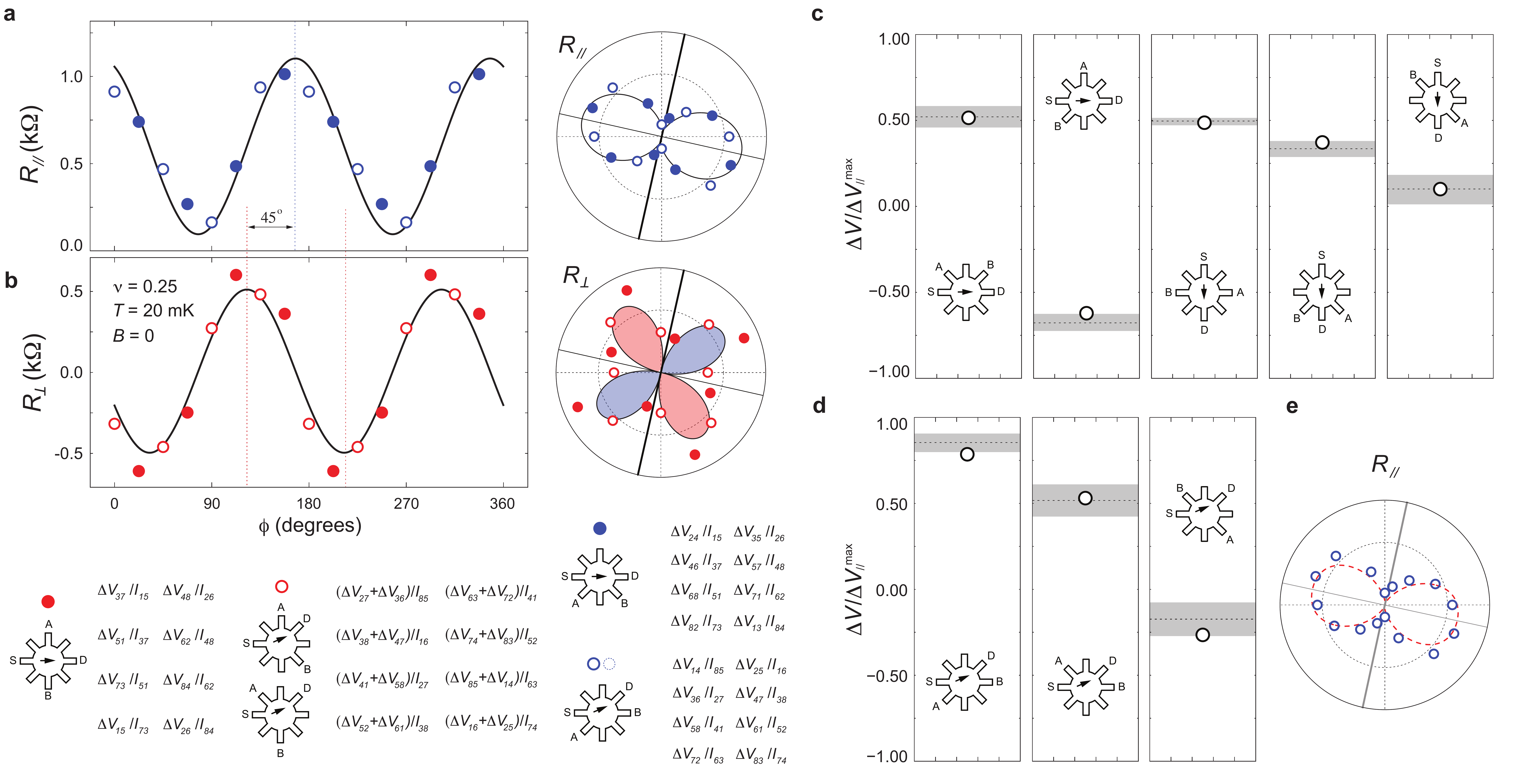}
\caption{\label{fig3} {\bf{Angle-resolved transport response of an anisotropic state at $T =20$ mK.}} (a) $R_{\parallel}$ and (b) $R_{\perp}$ as a function of azimuthal angle $\phi$, which denotes the direction of current flow. The right panels show polar coordinate plot the the same data. 
The measurement configurations for \Rpara\ and \Rperp\ are labeled in the legend. 
(c-d) $\Delta V$ measured using different configurations. The plotted value  is renormalized by the expected value of $\Delta V_{\parallel}$ when current flows along the anisotropy director, $\Delta V_{\parallel}^\textrm{max}$. The grey shaded horizontal stripes denote the expected range of voltage drop for each configuration, which is extracted from the anisotropic conductivity tensor and the boundary condition (see Fig.~\ref{figError} for more details) ~\cite{Vafek2022sunflower}.  (e) Based on the anisotropy conductivity tensor extracted from the combination of $56$ measurement configurations (panel (c-d) and Fig.~\ref{AnisotropicAll}), we calculate an expected angular dependence for \Rpara\ (red dash line), which is in excellent agreement with the measured $R_{\parallel}$ as a function of $\phi$ (blue circles). Grey solid lines indicate the principle axes of the anisotropy conductivity tensor. 
}
\end{figure*}

\Rpara\ and \Rperp\ only account for a fraction of $840$ possible measurement configurations available to the ``sunflower'' geometry. By measuring $\Delta V$ across all possible combinations of contacts, we can map the distribution of electrical potential along the circumference of the sunflower-shaped sample.
According to a recent calculation ~\cite{Vafek2022sunflower}, the potential distribution across a uniform sample is fully determined by the combination of conductivity tensor and the boundary condition, which is defined by the pair of contacts used for applying current bias (Fig.~\ref{fig1}e-h). As such, measuring $\Delta V$ across a number of different configurations allows us to simultaneously extract the conductivity tensor of the underlying electronic state and characterize the spatial uniformity across the sample.

We begin by analyzing the angle-resolved transport response at high temperature $T = 35$ K near the charge neutrality point (CNP) at $\nu=0.2$. Fig.~\ref{fig2}a-b shows a lack of angle-dependence for both \Rpara\ and \Rperp. At the same time, the value of \Rperp\ is close to zero for all azimuthal angles $\phi$, in stark contrast with the large value of \Rpara. Such angular dependence is in agreement with an isotropic state.
Moreover, we compare $\Delta V$  measured across $30$ configurations to the expected potential distribution of an isotropic state (horizontal stripes in Fig.~\ref{fig2}c). As shown in Fig.~\ref{fig2}c, the measured values for all $30$ configurations fall within the expected range of an isotropic state.  Since the model assumes a uniform sample, the excellent agreement with measurement points towards a uniform sample that is free of twist-angle inhomogeneity.

Starting from the isotropic state in Fig.~\ref{fig2},  an anisotropic state emerges with decreasing temperature. As shown in Fig.~\ref{fig3}a-b, the angular dependence of \Rpara\ and \Rperp\ both exhibit well-defined two-fold oscillation, which can be fit with the expected behavior of orthorhombic anisotropy ~\cite{Wu2017nematic,Wu2020nematic}, 
\begin{eqnarray}
R_{\parallel}(\phi) = \Delta R cos[2(\phi-\alpha-90^{\circ})]+R_0 \nonumber \\
R_{\perp}(\phi) = \Delta R cos[2(\phi-\alpha-45^{\circ})].
\end{eqnarray}
Here  $\Delta R$ denotes the oscillation amplitude and $R_0$ the average value of \Rpara$(\phi)$. The ratio between $\Delta R$ and $R_0$, \DRR, provides a measure of the electron anisotropy. $\alpha$ defines the orientation of the anisotropy director, which is a unit vector aligned along the principle axis with higher conductivity. The best fit to the angular dependence in Fig.~\ref{fig3}a-b yields $\alpha = 77^{\circ}$ and \DRR $=0.84$. This corresponds to a conductivity tensor with principle axis along $\phi = 77^{\circ}$ and $167^{\circ}$, which are marked by solid black lines in the polar coordinate plots in Fig.~\ref{fig3}a-b. When current flows along the principle axes, \Rpara\ is either maximized or minimized, whereas \Rperp\ vanishes. This accounts for the $45^{\circ}$ shift in the phase of the oscillation between \Rpara\ and \Rperp. The diagonal terms of the conductivity tensor, which denote sample conductivity along principle axes, are defined as \sigbar\ $+$ \dsig\ and \sigbar\ $-$ \dsig. According to the angle dependence in Fig.~\ref{fig3}a-b, the ratio between \dsig\ and \sigbar\ corresponds to $\delta \sigma / \bar{\sigma} =$ \DRR $=0.84$. This indicates a highly anisotropic electron state.

\begin{figure*}[!t]
\includegraphics[width=1\linewidth]{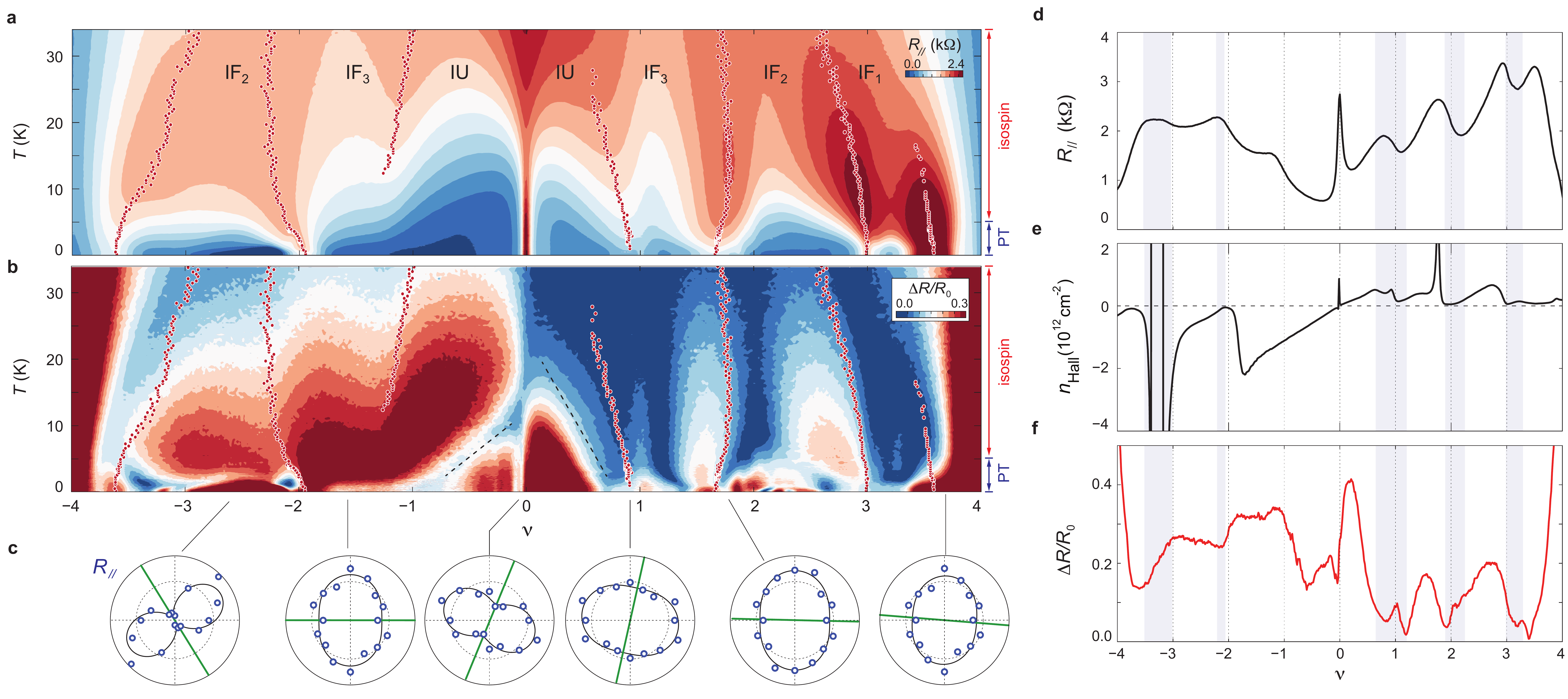}
\caption{\label{figN} {\bf{Electron anisotropy and cascade of isospin transitions.}}  $\nu-T$ map of (a) \Rpara\ and (b) anisotropy ratio \DRR. Boundaries between different isospin orders are marked by open white circles. (c) Polar coordinate plot of \Rpara\ measured at different moir\'e band fillings, showing prominent rotation in the anisotropy director (marked by green solid line).  (d) \Rpara, (e) Hall density $n_{Hall}$, and (f) \DRR\ as a function of moir\'e filling. Panel (d) and (f) are measured at $T = 10$ K, where electron anisotropy is dominated by the influence of the cascade phenomenon. }
\end{figure*}

\begin{figure*}
\includegraphics[width=0.9\linewidth]{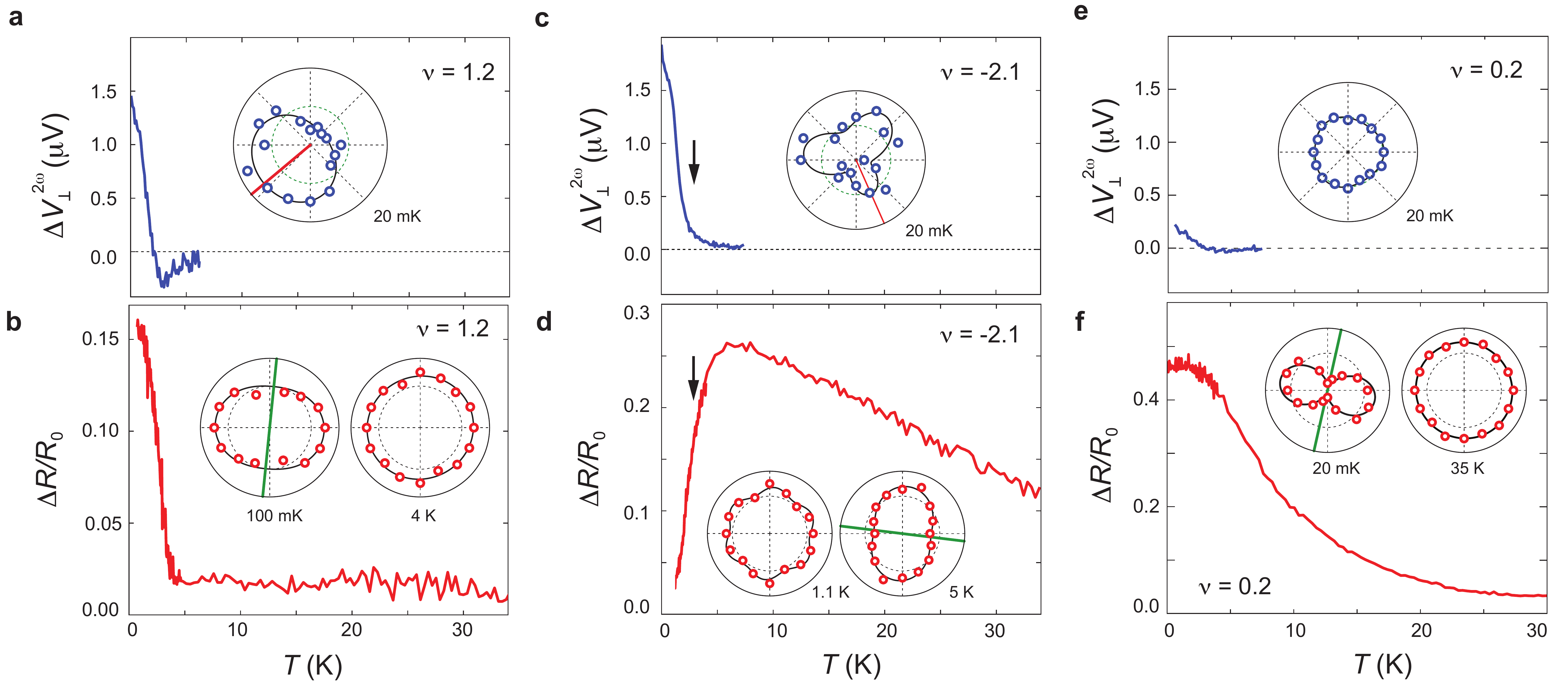}
\caption{\label{figT} {\bf{Interplay with the $PT$-breaking order.}} The temperature dependence of the second-harmonic nonlinear response $V^{2\omega}_{\perp}$ (panel (a), (c), and (e)) and the anisotropy ratio \DRR\ ((panel (b), (d), and (f))). The measurement is performed at band filling (a-b) $\nu=1.2$, (c-d) $\nu=-2.1$ and (e-f) $\nu=0.2$. The inset shows the polar-coordinate plot for the angular dependence of \Rpara\ and $V^{2\omega}_{\perp}$. The temperature dependence of the anisotropy ratio at low temperature, which describes the strength of electron anisotropy in the linear transport response, is dependent on the angular symmetry and strength of the $PT$-breaking order, which is manifested by the angular dependence of the nonlinear transport response at second-harmonic frequency. }
\end{figure*}

At this doping and temperature, mapping the potential distribution across the ``sunflower'' sample testifies that the entire sample is described by the same anisotropic conductivity tensor. Fig.~\ref{AnisotropicAll} plots the voltage difference of more than $50$ measurement configurations. Collectively, these measurements are best fit with a single conductivity matrix. The quality of this fit is demonstrated by the excellent agreement between the measurement and the expected value from the calculated potential distribution (horizontal stripes in Fig.~\ref{fig3}c-d and Fig.~\ref{AnisotropicAll}). Most importantly, this fit produces an anisotropy director along $\alpha = 78^{\circ}$ and an anisotropy ratio \DRR $=\delta \sigma / \bar{\sigma}=0.74$, which is in excellent agreement with the conductivity tensor extracted from in Fig.~\ref{fig3}a-b.  The consistency demonstrated by different schemes of ARTM offers further validation for the identification of electronic anisotropy.

Having established the method of ARTM, we are now in position to examine the connection between the observed electron anisotropy and Coulomb interaction. 
In graphene-based moir\'e systems, strong Coulomb interaction drives a cascade of isospin transitions. This gives rise to a unique doping-dependent modulation in the transport response. For instance, Fig.~\ref{figN}a shows the $\nu-T$ map of \Rpara. The isospin transitions divide the moir\'e flatband into regimes of different isospin orders, with the boundary defined by peak positions of \Rpara, along with resets in Hall density (see Fig.~\ref{figCascade}) ~\cite{Rozen2020pomeranchuk,Saito2021pomeranchuk,Park2021flavour,Liu2022DtTLG}. We mark the isospin order of each regime, such as isospin ferromagnet IF and isospin unpolarized IU, which are identified based on the main sequence of quantum oscillation (see Fig.~\ref{figSIfan}).
The cascade of isospin transitions, which occur near most integer band fillings (with fully filled/empty moir\'e band defined as band filling $\pm4$), also coincide with resets in the Hall density (Fig.~\ref{figCascade})  ~\cite{Rozen2020pomeranchuk,Saito2021pomeranchuk,Park2021flavour,Liu2022DtTLG}. The presence of cascade phenomenon provides a unique window allowing us to characterize the link between Coulomb interaction and electron anisotropy.
This is achieved by measuring angle-resolved transport response across the $\nu-T$ map. As shown in Fig.~\ref{figPolar}, the conductivity tensor, including the anisotropy ratio \DRR\ and director orientation $\alpha$, can be extracted by fitting \Rpara\ and \Rperp\ using Eq.~1. Across the moir\'e flatband, both \DRR\ and $\alpha$ display prominent dependence on moir\'e band filling. Most importantly, the electronic state is shown to be more (less) anisotropic at low (high) temperature, which provides a strong indication that the spatial anisotropy is an emergent phenomenon (Fig.~\ref{figPolar}). In the following, we will examine the interplay between electron anisotropy, cascade phenomenon and other electronic orders across the moir\'e flatband by plotting the anisotropy ratio \DRR\ across the $\nu-T$ map in Fig.~\ref{figN}b. The director orientation $\alpha$, which is extracted by fitting the same angular dependence, is shown in Fig.~\ref{figAlpha}c.

First, we examine the evolution of electron anisotropy in the temperature range of $5 < T < 35$ K, where the the cascade phenomenon dominates.
The doping-dependence of anisotropy ratio \DRR\ is shown to be in excellent correspondence with the cascade phenomenon across the $\nu-T$ map. Near each isospin transition (marked as open white circles in Fig.~\ref{figN}a-b), we observe a local maximum and minimum in the anisotropy ratio, which are located on either side of the transition.  
This correlation is further demonstrated by examining the doping dependence measured at a fixed temperature $T = 10$ K. As shown in Fig.~\ref{figN}e, the reset in Hall density $n_{Hall}$ gives rise to a small Fermi surface on the high density side of the isospin transition. In these density regimes (marked with blue shaded stripes), the transport response is mostly independent of the azimuthal direction of current flow, which points towards an isotropic electron state. On the other hand, prominent electron anisotropy, evidenced by strong angular dependence in the transport response, is associated with the large Fermi surface on the low density side of the isospin transition. 
The anisotropic director (marked by green solid lines) exhibits prominent doping-dependent rotation, as shown in Fig.~\ref{figN}c, which is comparable with previous observations in cuprate ~\cite{Wu2017nematic} and graphene-based moir\'e systems ~\cite{Xie2019STM,Jiang2019STM,Choi2019STM,Samajdar2021nematic}.

The direct link between electron anisotropy and Coulomb-driven isospin transition is further confirmed by analyzing the twist angle dependence.
When the twist angle is detuned from the magic-angle, the moir\'e band structure becomes more dispersive, diminishing the influence of Coulomb interaction ~\cite{DWPaper}. As a result, the abundance of isospin transitions near the magic angle is reduced to a single Fermi surface reconstruction near $\nu=+2$ at $\theta = 1.33^{\circ}$. This is evidenced by the Hall density reset marked by the blue shaded stripe in Fig.~\ref{figSmall}a-b. In the absence of isospin transition, electron anisotropy is suppressed. This is especially the case in the hole-doping band, where vanishing anisotropy ratio points towards an isotropic electron state (Fig.~\ref{figSmall}a). 
Together, our findings provide unambiguous evidence that Coulomb-driven cascade phenomenon plays an essential role in the doping and temperature dependence in electron anisotropy.

Notably, the behavior of electron anisotropy in the temperature range of $T < 5$ K deviates from the cascade phenomenon.
In this temperature range, the moir\'e flatband of twisted trilayer graphene hosts a novel electronic order that breaks both parity and time-reversal symmetry ($PT$-breaking), which is evidenced by the angular dependence of the nonreciprocal transport response ~\cite{Zhang2022diodic}. Notably, the emergence of this $PT$-breaking order at $T < 5$ K shows excellent correspondence with the temperature dependence of electron anisotropy. 
When the low-temperature nonreciprocal response is one-fold symmetric (Fig.~\ref{figT}a), the onset of nonreciprocity coincides with the enhancement in electron anisotropy, which is evidenced by the sharp onset in the anisotropy ratio (Fig.~\ref{figT}b). On the contrary, a predominantly three-fold symmetric nonreciprocal response, as shown in the inset of Fig.~\ref{figT}c, induces a suppression in the anisotropy ratio with decreasing temperature (Fig.~\ref{figT}d). This gives rise to an isotropic state at low temperature (left inset in Fig.~\ref{figT}d), even though the high temperature state is highly anisotropic (right inset in Fig.~\ref{figT}d). Similar low-temperature behavior in electron anisotropy is observed at a number of moir\'e band fillings, as shown in Fig.~\ref{figN}b and Fig.~\ref{figSIT}). 
Near the CNP, a small nonreciprocal response at $\nu=0.2$ points towards a weak $PT$-breaking order (Fig.~\ref{figT}e). At this band filling, the temperature dependence of the anisotropy ratio exhibits no sharp changes at $T < 5$ K (Fig.~\ref{figT}f). That the temperature dependence of electron anisotropy is determined by the angular symmetry, as well as magnitude, of nonreciprocity points towards the dominating influence of the $PT$-breaking order at low temperature.   Combined, our findings suggest that changes in electron anisotropy at $T < 5$ K originates from the emergence of the $PT$-breaking order. Owing to the time-reversal breaking, the $PT$-breaking order does not couple to lattice distortion. Therefore, the associated electron anisotropy must have a Coulomb origin. This provides another indication for a direct link between electron anisotropy and Coulomb interaction.

In a realistic solid state sample, some level of lattice distortion is unavoidable. The influence of uniaxial strain is evidenced in our ARTM as well. For instance, the onset of electron anisotropy in Fig.~\ref{figT}f is distributed over a wide temperature window.  A broadened transition could be the result of uniaxial strain in the sample. 
Nevertheless, the evolution of electron anisotropy as a function of moir\'e doping and twist angle demonstrates an unambiguous link with the cascade phenomenon and the $PT$-breaking order. Combined, our findings  point towards the crucial influence of Coulomb interaction in stabilizing electron anisotropy.

\section*{Acknowledgments}
N.J.Z. acknowledge support from the Jun-Qi fellowship. J.I.A.L. acknowledge funding from NSF DMR-2143384. Device fabrication was performed in the Institute for Molecular and Nanoscale Innovation at Brown University.
K.W. and T.T. acknowledge support from the Elemental Strategy Initiative
conducted by the MEXT, Japan (Grant Number JPMXP0112101001) and  JSPS
KAKENHI (Grant Numbers 19H05790, 20H00354 and 21H05233). O. V. was supported by NSF Grant No. DMR-1916958 and is partially funded by the Gordon and Betty Moore Foundation’s EPiQS Initiative Grant GBMF11070, National High Magnetic Field Laboratory through NSF Grant No. DMR-1157490 and the State of Florida.

\bibliography{Li_ref}

\newpage

\newpage
\clearpage

\pagebreak
\begin{widetext}
\section{Supplementary Materials}

\begin{center}
\textbf{\large Electronic anisotropy in magic-angle twisted trilayer graphene}\\
\vspace{10pt}

Naiyuan J. Zhang,
Yibang Wang, K. Watanabe, T. Taniguchi, and J.I.A. Li$^{\dag}$

\vspace{10pt}
$^{\dag}$ Corresponding author. Email: jia$\_$li@brown.edu
\end{center}

\noindent\textbf{This PDF file includes:}

\noindent{Supplementary Text}

\noindent{Materials and Methods}

\noindent{Figs. S1 to S10}

\renewcommand{\vec}[1]{\boldsymbol{#1}}

\renewcommand{\thefigure}{S\arabic{figure}}
\def\theequation{S\arabic{equation}}
\def\thetable{S\Roman{table}}
\setcounter{figure}{0}
\setcounter{equation}{0}

\newpage

\begin{figure*}
\includegraphics[width=1\linewidth]{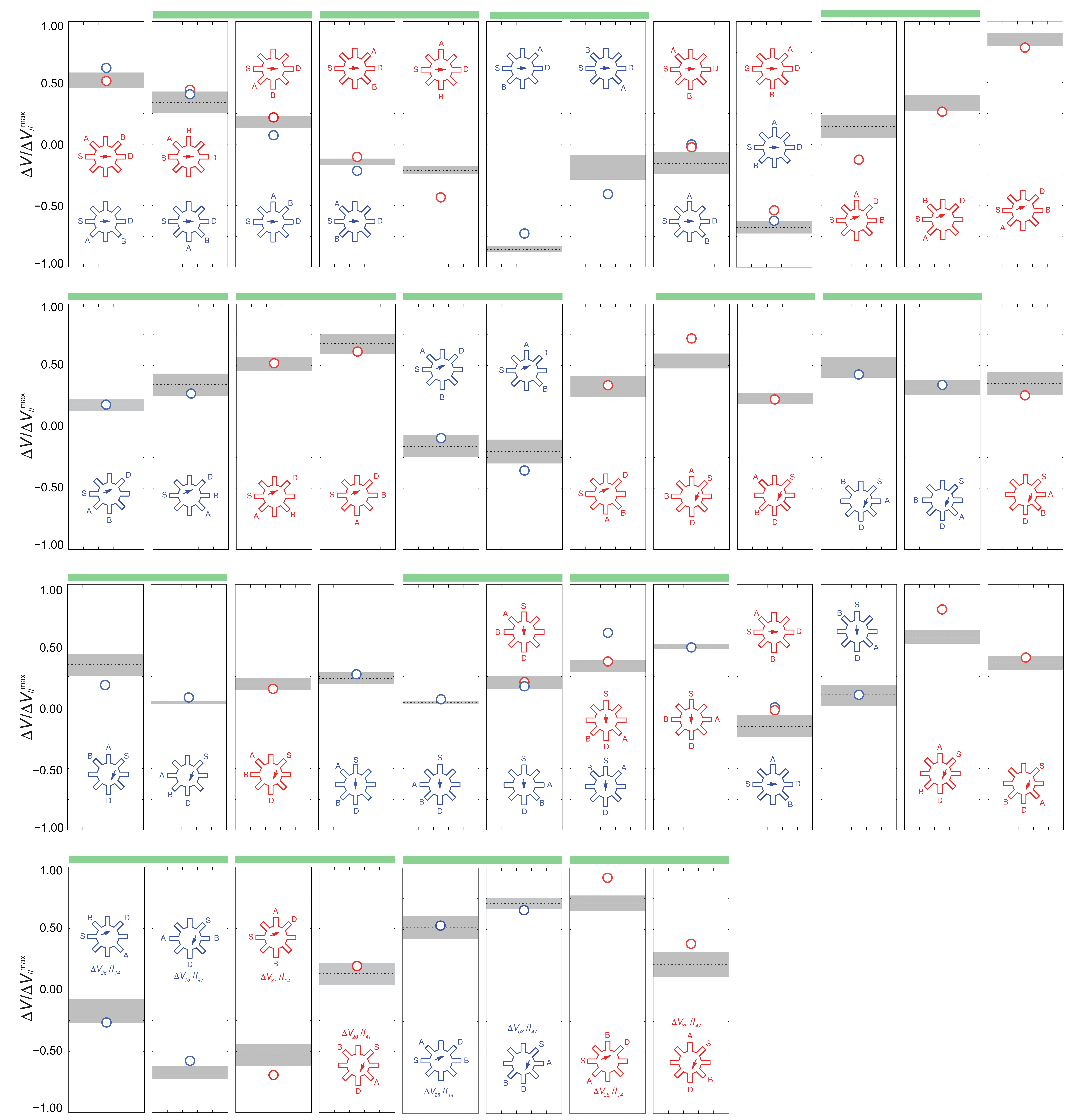}
\caption{\label{AnisotropicAll} {\bf{ARTM on a highly anisotropic state.}} $\Delta V$ measured from more than $50$ different configurations.
An anisotropy conductivity tensor is extracted by fitting all measurement results with the model discussed in Ref.~\cite{Vafek2022sunflower}.
The grey shaded horizontal stripes denote the expected range of voltage drop for each configuration, calculated using the anisotropic conductivity tensor and the boundary condition ~\cite{Vafek2022sunflower}. 
The plotted values of $\Delta V$ are renormalized by the expected value of $\Delta V_{\parallel}$ when current flows along the anisotropy director, $\Delta V_{\parallel}^\textrm{max}$. Green bars indicate that highlighted configurations share the same expected value for an isotropic conductivity tensor.  }
\end{figure*}


\begin{figure*}
\includegraphics[width=0.65\linewidth]{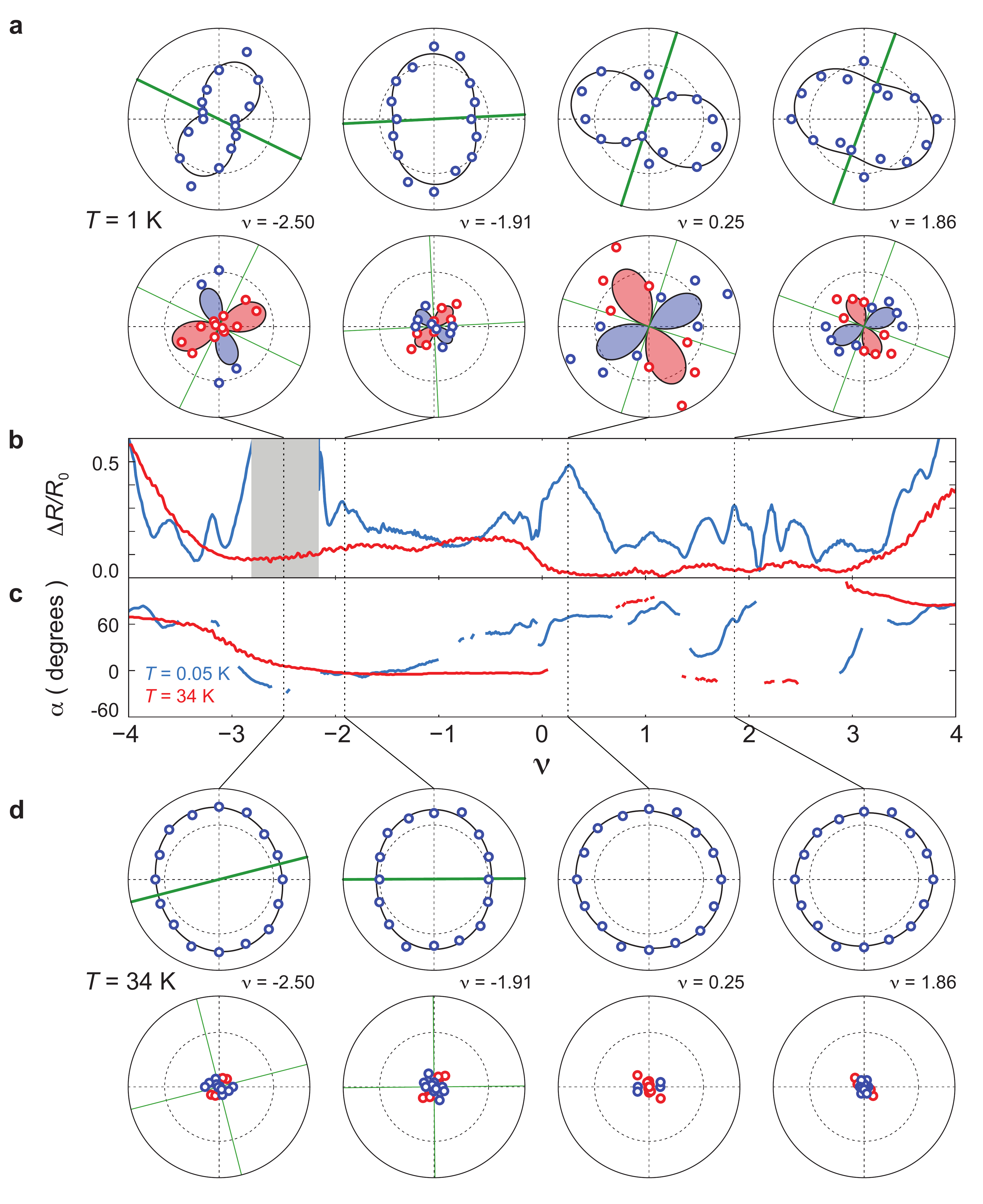}
\caption{\label{figPolar} {\bf{Polar-coordinate plot of \Rpara\ and \Rperp\ measured at different temperatures.} }  
(a) Polar-coordinate plot of \Rpara\ and \Rperp\ measured at $T = 1$ K and different band fillings. (b) Anisotropy ratio and (c) director orientation as a function of moir\'e band filling $\nu$, measured at different temperatures. (d) Polar-coordinate plot of \Rpara\ and \Rperp\ measured at $T = 34$ K and different band fillings. Electron anisotropy is much more prominent at low temperature, whereas transport measurement at $34$ K points towards weakly anisotropic and isotropic angular dependence across the entire moir\'e band. At the same time, the director orientation $\alpha$ shows more prominent doping-induced variation at $T = 1$ K. The temperature dependence suggests that electron anisotropy is an emergent phenomenon that is associated with the electronic order.
}
\end{figure*}

\begin{figure*}
\includegraphics[width=0.95\linewidth]{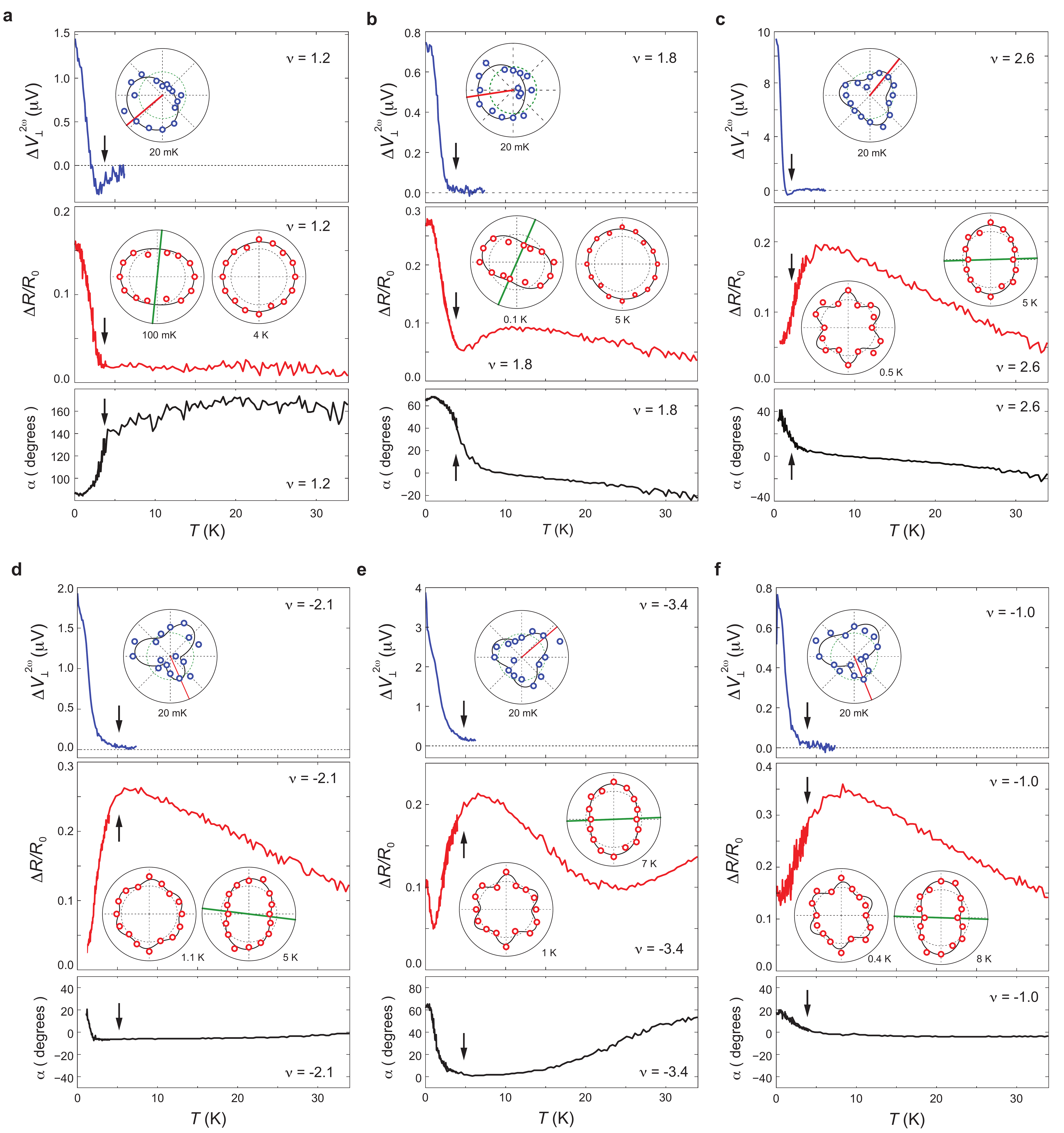}
\caption{\label{figSIT} {\bf{The connection between the loop current state and orthorhombic anisotropy.} }  
The temperature dependence of \Vto\ (top panels), \DRR\ (middle panels), and $\alpha$ (bottom panels) as a function of temperature measured at (a) $\nu=1.2$, (a) $\nu=1.8$, (c) $\nu=2.6$,  (d) $\nu=-2.1$, (e) $\nu=-3.4$, and (f) $\nu=-1.0$. \Vto\  is the nonlinear response measured at the second-harmonic frequency with an AC current of $I_{AC} = 100$ nA, whereas \DRR\ is extracted from the angular dependence of \Rpara, which is the linear transport response measured with a DC current of $I_{DC} = 5$ nA. The inset shows polar-coordinate plot of \Vto\ and \Rpara\ measured at various moir\'e band fillings and temperatures. In (a-b), the angular dependence of the nonreciprocal response is one-fold symmetric. The onset of \Vto\ coincides with an enhancement in the anisotropy ratio of the linear transport response. In (c-f), \Vto\ is predominantly three-fold symmetric. The onset of \Vto\ coincides with a decrease in the anisotropy ratio of the linear transport response. The angular dependence of the linear transport response at low temperature preserves three-fold rotational symmetry $C_3$ as well. 
}
\end{figure*}

\begin{figure*}
\includegraphics[width=0.78\linewidth]{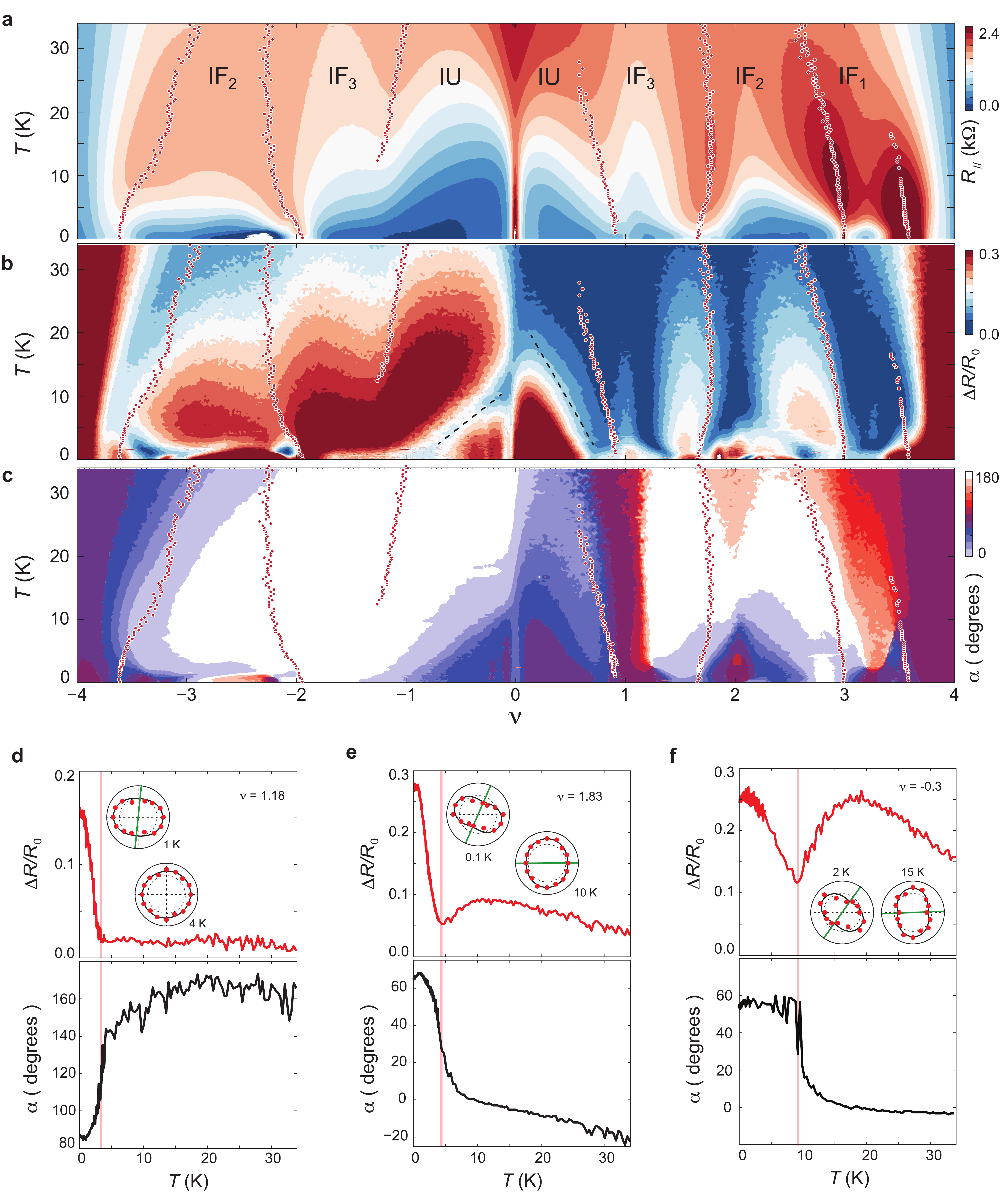}
\caption{\label{figAlpha} {\bf{Director orientation.} }  
$\nu-T$ map of (a) \Rpara, (b) anisotropy ratio \DRR, and (c) director orientation $\alpha$. (d-f) Temperature dependence of the anisotropy ratio (top axis) and director orientation (bottom panel) measured at different band fillings. According to Fig.~\ref{figT} and Fig.~\ref{figSIT}, the onset in \DRR\ with decreasing temperature corresponds to the emergence of a $PT$-breaking electronic order. That this onset coincides with a rotation in the anisotropy director (marked with green solid line in the polar-coordinate plot) suggests that director orientation can be used to identify changes in the electronic order. Panel (f) plots the behavior of electron anisotropy across the boundary defined by black dashed line in panel (b). The anisotropy ratio displays a minimum at the boundary, concomitant with a jump in the director orientation. This points towards a transition between different electronic orders across the black dashed line. Notably, while this transition corresponds to clear features in the anisotropy ratio and director orientation, it is not visible in conventional transport response (Fig.~\ref{figN}a and panel (a)). 
}
\end{figure*}

\begin{figure*}
\includegraphics[width=0.9\linewidth]{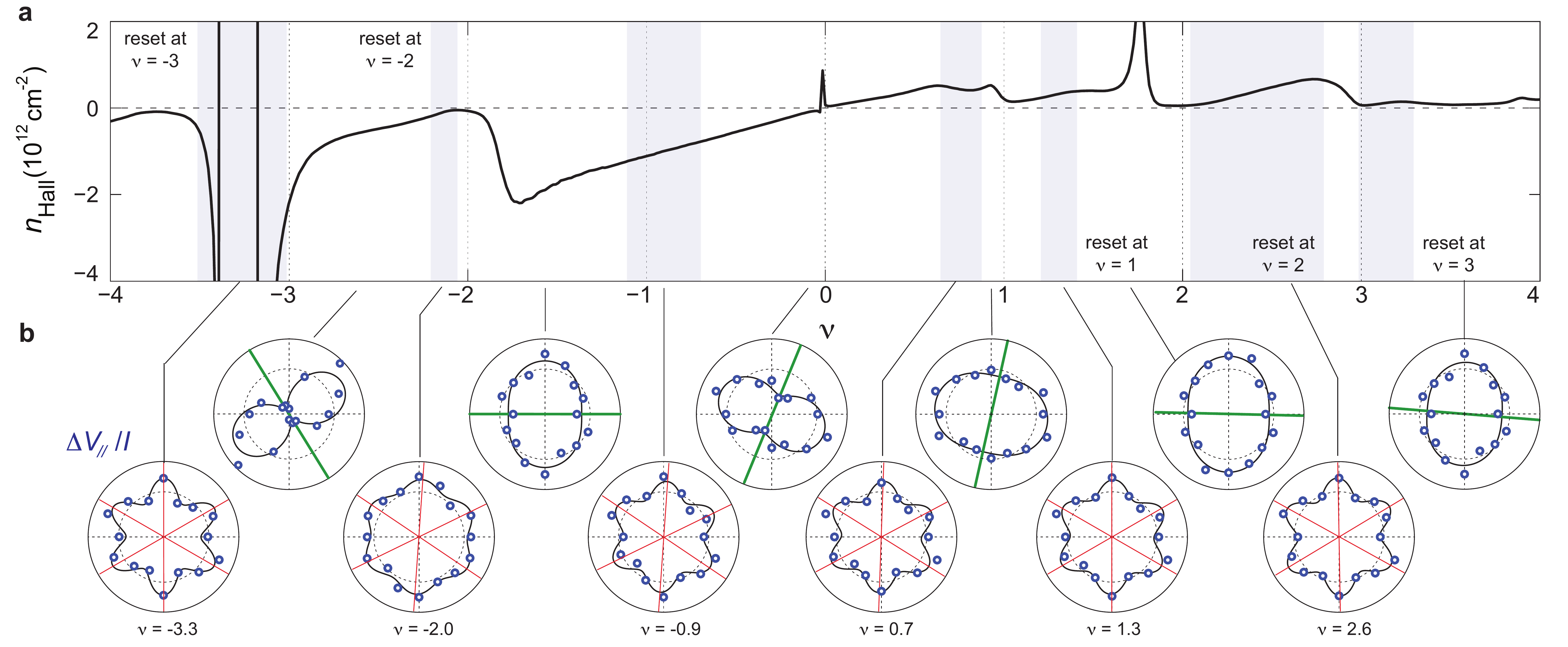}
\caption{\label{fig6fold} {\bf{$C_3$-preserving and $C_3$-breaking orders.} }  
At low temperature, $T = 20$ mK, angle-resolved transport response alternates between $C_3$-preserving and $C_3$-breaking behaviors. $C_3$-preserving response is observed in density regimes marked by blue shaded stripes in panel (a), which are located near integer band fillings. $C_3$-breaking response, which corresponds to orthorhombic anisotropy, is predominantly observed away from the integer filling. This density dependence is consistent with the $\nu-T$ map in Fig.~\ref{figN}b. Notably, the $C_3$-preserving response shows maximum and minimum \Rpara\ along the same azimuth angles across the entire moir\'e flatband. These angles are in excellent agreement with the three-fold symmetric nonreciprocal response associated with the $PT$-breaking order ~\cite{Zhang2022diodic}. This implies that the  $C_3$-preserving response is directly linked to the shape of the underlying Fermi surface. Our angle-resolved measurements in Fig.~\ref{fig2} and Fig.~\ref{fig3} indicate a uniform response across the entire sample. \emph{i.e.}, the potential influence of anisotropy domain of secondary importance. As such, the $C_3$-preserving response cannot be explained by domains with different anisotropy directors.
}
\end{figure*}

\begin{figure*}
\includegraphics[width=0.9\linewidth]{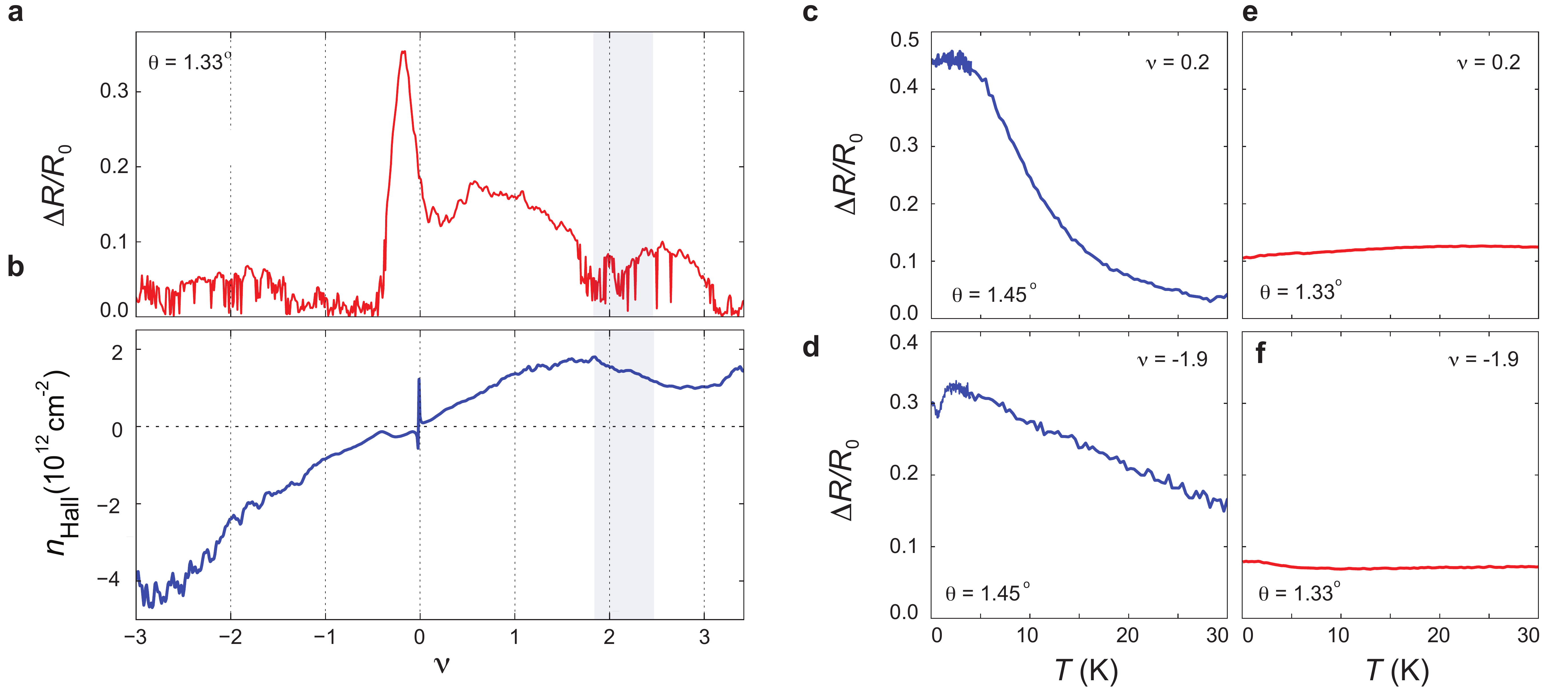}
\caption{\label{figSmall} {\bf{Twist angle dependence.}} (a) Anisotropy ratio \DRR\ and (b) Hall density $n_{Hall}$ as a function of moir\'e band filling measured in a sample with twist angle $\theta = 1.33^{\circ}$. (c-f)  Anisotropy ratio \DRR\ as a function of temperature measured at different moir\'e band fillings in a sample (c-d) near the magic angle, and (e-f) with twist angle  $\theta = 1.33^{\circ}$. }
\end{figure*}

\clearpage
\newpage

\section{Materials and Method}

\subsection{Device Fabrication}

The doubly encapsulated \tlg\ is assembled using the “cut-and-stack” technique. All components of the structure are assembled from top to bottom using the same poly(bisphenol A carbonate) (PC)/polydimethylsiloxane (PDMS) stamp mounted on a glass slide. The sequence of stacking is: graphite as top gate electrode, 24 nm thick hBN as top dielectric, bilayer \WSe, \tlg, 24 nm thick hBN as bottom dielectric, bottom graphite as bottom gate electrode. The entire structure is deposited onto a doped Si/SiO2 substrate. Electrical contacts to tTLG are made by CHF$_3$/O$_2$ etching and deposition of the Cr/Au (2/100 nm) metal edge contacts. The sample is shaped into an sunflower geometry with an inner radius of 1.9 $\mu$m for the circular part of the sample. In this geometry, the electrical contacts are separated by an azimuth angle of $45^o$, allowing an increment in the azimuth angle that is  $22.5^o$. 

\subsection{Transport measurement}

The carrier density in tTLG is tuned by applying a DC voltage bias to the bottom gate electrode. The electrical potential of the top gate electrode is held at zero. As a result, the tTLG sample experience a non-zero displacement field $D$ at large carrier density, which induces hybridization between the monolayer band and the moir\'e flatband. We note that the dependence of Hall density on moir\'e band filling is in excellent agreement with $D=0$ behavior from previous observations. This indicates that the influence of $D$ on the moir\'e flatband is not substantial. This is further confirmed by the Landau fan diagram in Fig.~\ref{figSIfan}, which is also consistent with the expected behavior at $D=0$. 

Transport measurement is performed in a BlueFors LD400 dilution refrigerator with a base temperature of 20 mK. Temperature is measured using a resistance thermometer located on the cold finger connecting the mixing chamber and the sample. An external multi-stage low-pass filter is installed on the mixing chamber of the dilution unit. The filter contains two filter banks, one with RC circuits and one with LC circuits. The radio frequency low-pass filter bank (RF) attenuates above 80 MHz, whereas the low frequency low-pass filter bank (RC) attenuates from 50 kHz. The filter is commercially available from QDevil. 

The current-voltage characteristics is measured using two methods. In the DC measurements, we sweep the amplitude of the DC current with a small, fixed AC excitation of $5$ nA at a frequency of $13$ Hz. The differential voltage is measured using standard lock-in techniques with Stanford Research SR830 amplifier. In the AC measurements, we sweep the amplitude of the AC current at a frequency of $13$ Hz. The nonlinear response is measured at the second harmonic frequency using Stanford Research SR830 amplifier. 

Transport response is measured across voltage leads that are parallel and perpendicular to the current flow direction. The setup for the parallel response, $\Delta V_{\parallel}$, is shown in Fig.~\ref{fig1}a. For current flowing in the azimuth angle $0-180^{\circ}$, Fig.~\ref{figMeasurement}a-h display $8$ measurement configurations with an increment of $22.5^{\circ}$ in the azimuth direction of current flow. The voltage measurement in panel e-h is different by a geometric factor compared to that of panel a-d. This geometric factor is shown to be $1.09$ (Fig.~\ref{fig2}).

\begin{figure*}
\includegraphics[width=0.57\linewidth]{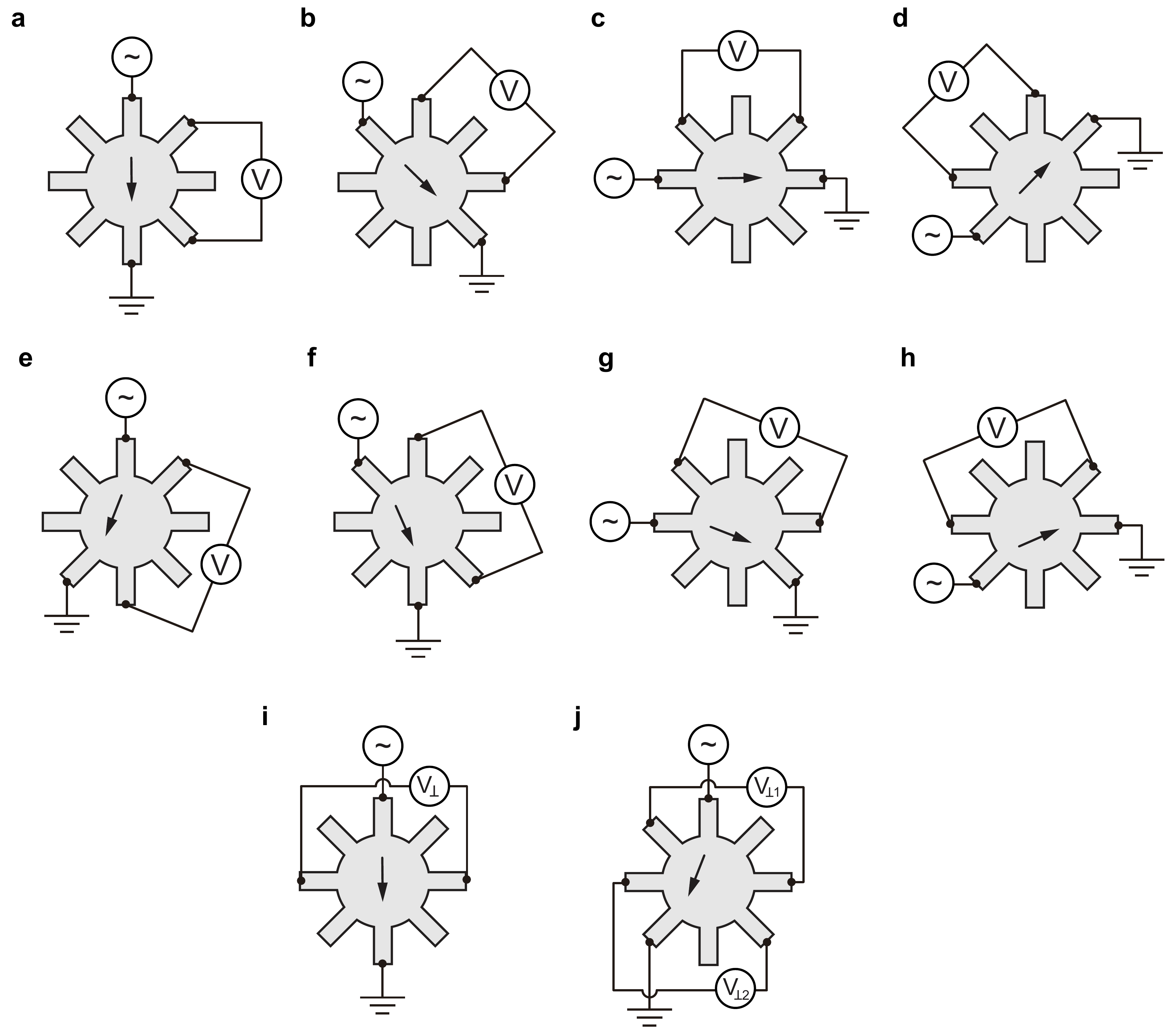}
\caption{\label{figMeasurement} {\bf{Schematic of angle-resolved transport measurement.}} (a-h) Eight measurement configurations for \Rpara $=\Delta V_{\parallel}/I$, which corresponds to either different current flow directions. With forward and reverse DC bias, this gives us 16 angles. Measurement configuration for $\Delta V_{\perp}$ in (i) configuration I and (j) configurations II. In configuration I, $\Delta V_{\perp}=0$ for an isotropic state (panel $iii$ in Fig.~\ref{fig2}). In configuration II, while both $\Delta V_{\perp 1}$ and $\Delta V_{\perp 2}$ are non-zero for an isotropic state, $\Delta V_{\perp 1}+\Delta V_{\perp 2} = 0$, as shown in panel $iv$ and $v$ in Fig.~\ref{fig2}. As such, we define \Rperp\ as \Rperp $=(\Delta V_{\perp 1}+\Delta V_{\perp 2})/I$ for configuration II, and \Rperp $=\Delta V_{\perp}/I$ for configuration I.
}
\end{figure*}

\begin{figure*}
\includegraphics[width=0.57\linewidth]{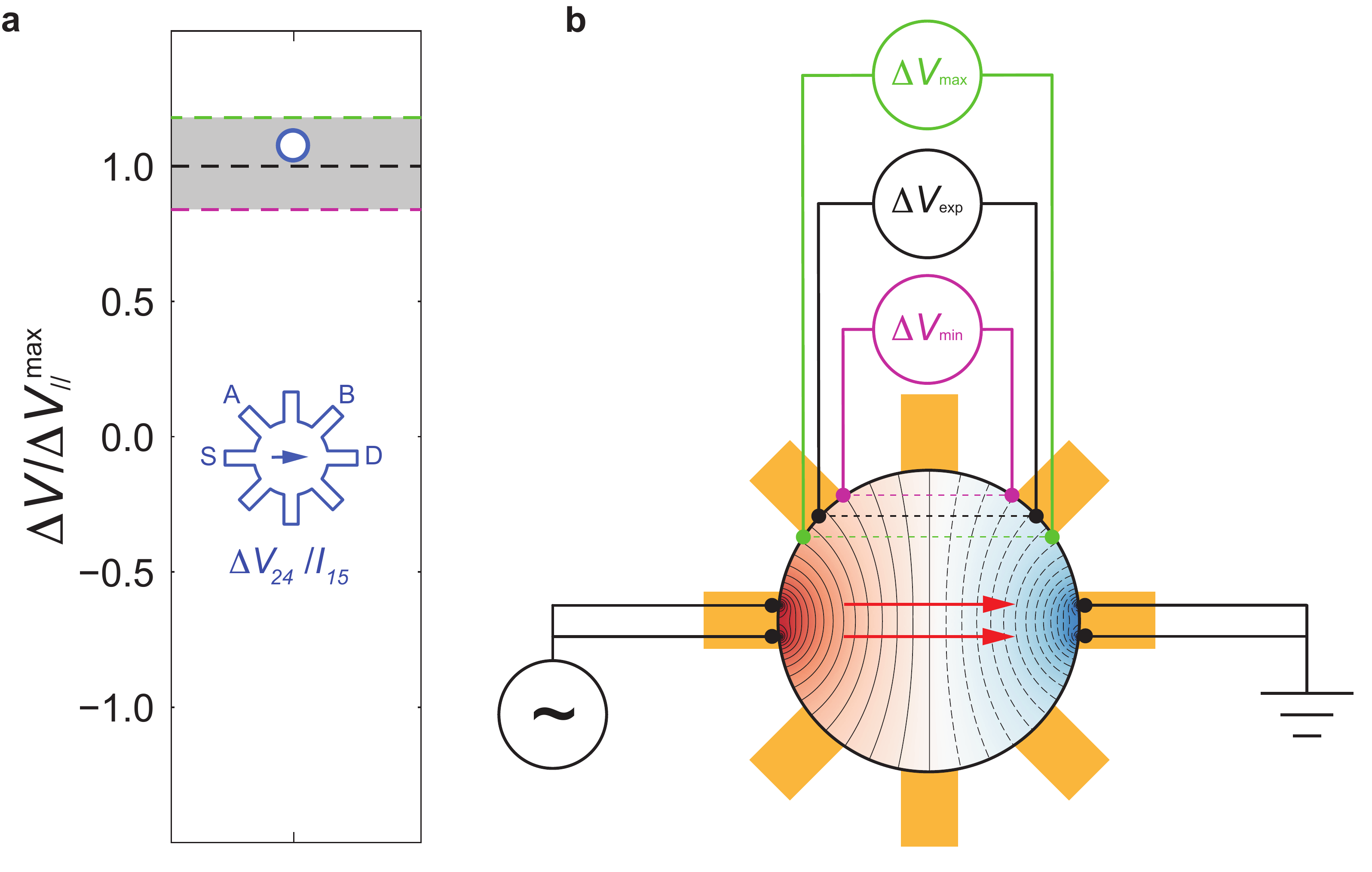}
\caption{\label{figError} {\bf{Expected range of distribution for $\Delta V$.}} (a) The open blue circle denote the measured value of $\Delta V$ for a specific configuration, $\Delta V_{24}/I_{15}$. The grey horizontal stripe marks the expected range of distribution based on the potential distribution. As shown in panel (b), the range of distribution for $\Delta V$ arise from the non-zero width of electrical contact. Based on the potential distribution across the sample ~\cite{Vafek2022sunflower}, the range of $\Delta V$ across a pair a contact is defined by the maximum and minimum potential difference, which are measured between green dots (maximum potential difference) and purple dots (minimum potential difference). These correspond to the green and purple dashed lines in panel (a). Notably, we have also taken into account the non-zero width of the current bias contact. Although the influence of the current bias contact appears to be of secondary importance. 
}
\end{figure*}

\subsection{The cascade phenomenon}

The $\nu-T$ map of the moir\'e flatband is divided into different areas based on the underlying isospin polarization. The boundaries of different isospin polarizations are defined by peaks in longitudinal resistance, concomitant with reset in the Hall density ~\cite{Saito2021pomeranchuk,Rozen2020pomeranchuk,Liu2022DtTLG,Park2021flavour}. Fig.~\ref{figCascade} shows the $\nu-T$ map of $R_{\parallel}$ and Hall density $n_{Hall}$. Isospin transitions are marked by white circles in the top panel. The cascade of isospin transition is clearly detectable at $T > 30$ K. This onset of isospin polarization transitions provides a characteristic for the Coulomb energy scale, which is believed to be the main driver behind the observed cascade phenomenon. 

\begin{figure*}
\includegraphics[width=0.9\linewidth]{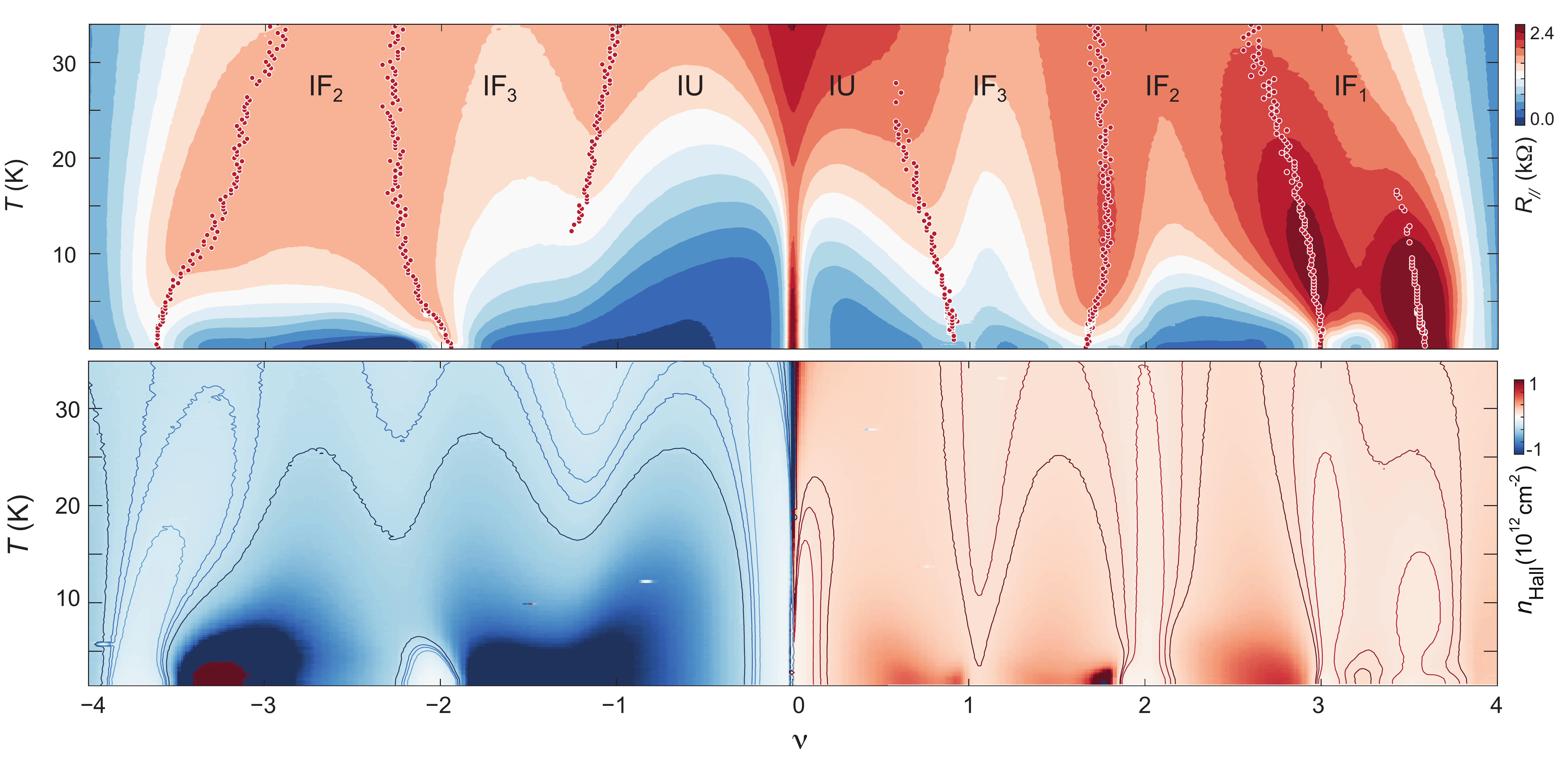}
\caption{\label{figCascade} {\bf{Cascade of isospin transitions.} }  
\Rpara\ (top panel) and Hall density $n_{Hall}$ (bottom panel) measured with linear transport (small current bias of $5$ nA) as a function of moir\'e filling and temperature. The peak position of $\Delta V_{\parallel}/I$ marks the boundary between different isospin orders ~\cite{Saito2021pomeranchuk,Liu2022DtTLG}, which are marked with white circles. The transition between different isospin orders coincide with resets in the Hall density. The cascade of isospin transitions are detectable at $T = 34$ K, much higher compared to the onset temperature of valley-polarized loop current state, which is $T < 5$ K.  
}
\end{figure*}

\begin{figure*}
\includegraphics[width=0.7\linewidth]{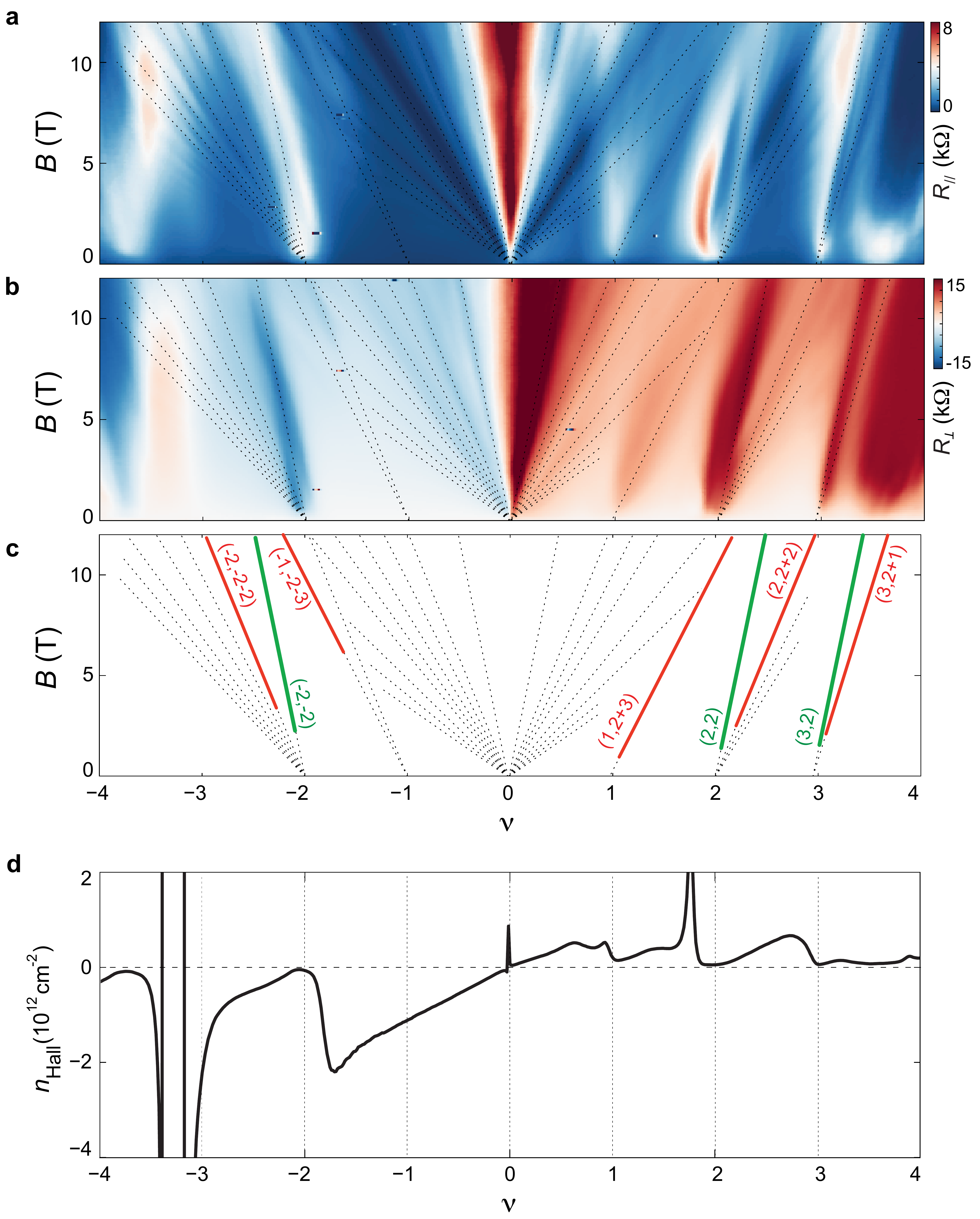}
\caption{\label{figSIfan} {\bf{Magneto-transport measurement across the moir\'e flatband.} }  
(a) \Rpara\ and (b) \Rperp\ across the moir\'e filling-magnetic field ($\nu-B$) map. Incompressible states are manifested as minima in $R_{\parallel}/I$, concomitant with quantized plateau in $R_{\perp}/I$. (c) The most prominent incompressible states are marked with black dashed line in the schematic $\nu-B$ map, where each trajectory is described by a pair of quantum numbers ($t, s$) from the Diophantine equation $\nu = t \phi/\phi_{0} + s$. Here $\nu$ is the moir\'e filling factor at the incompressible state, whereas $t$ and $s$ describe the slope and intercept of each trajectory ~\cite{Xie2021tblg,Spanton2018}.
(d) Hall density $n_{Hall}$ as a function of moir\'e filling measured at $B = 0.5$ T.
}
\end{figure*}

\end{widetext}
\end{document}